\documentclass[12pt]{iopart}

\usepackage{lineno}
\usepackage[T1]{fontenc}
\usepackage[integrals]{wasysym}
\usepackage{comment}
\usepackage{graphicx}

\newcommand{\bra}[1]{\ensuremath{\langle #1|}}
\newcommand{\ket}[1]{\ensuremath{|#1 \rangle}}
\newcommand{\calL}{\ensuremath{\mathcal{L}}}
\newcommand{\calM}{\ensuremath{\mathcal{M}}}
\newcommand{\p}{\ensuremath{\partial}}
\newcommand{\proj}[2]{\ensuremath{\langle #1 | #2 \rangle}}

\begin{document}
\title{Adaptive quantum accelerated imaging for space domain awareness}

\author{Hyunsoo Choi, Fanglin Bao, and Zubin Jacob$^{*}$}

\address{Purdue University, School of Electrical and Computer Engineering, Birck Nanotechnology Center, 47907, West Lafayette, Indiana, USA}

\ead{zjacob@purdue.edu}

\begin{abstract}
The growth in space activity has increased the need for Space Domain Awareness (SDA) to ensure safe space operations. Imaging and detecting space targets is, however, challenging due to their dim appearance, small angular size/separation, dense distribution, and atmospheric turbulence. These challenges render space targets in ground-based imaging observations as point-like objects in the sub-Rayleigh regime, with extreme brightness contrast but a low photon budget. Here, we propose to use the recently developed quantum-accelerated imaging (QAI) for the SDA challenge. We mainly focus on three SDA challenges (1) minimal a priori assumptions (2) many-object problem (3) extreme brightness ratio. We also present results on source estimation and localization in the presence of atmospheric turbulence. QAI shows significantly improved estimation in position, brightness, and number of targets for all SDA challenges. In particular, we demonstrate up to 2.5 times better performance in source detection than highly optimized direct imaging in extreme scenarios like stars with a 1000 times brightness ratio. With over 10,000 simulations, we verify the increased resolution of our approach compared to conventional state-of-the-art direct imaging paving the way towards quantum optics approaches for SDA.
\end{abstract}

\vspace{2pc}

%
%
%

\section{Introduction}

Over the past decades, space exploration and space missions have witnessed a remarkable surge, resulting in a significant accumulation of space objects such as satellites, spacecraft, and debris in Earth's orbit \cite{pak2018robust}. These objects pose substantial threats to both astronauts and equipment in current and future space endeavors, necessitating their detection and monitoring to ensure the safety and security of space operations. This essential task is referred to as Space Domain Awareness (SDA), which aims to continuously detect, track, catalog, and identify artificial objects in the space environment \cite{blake2012space}. However, various factors can prevent the effective detection of space objects, including dim appearance, small size, and a considerable distance from measuring instruments \cite{ye2015autonomous}. These challenges necessitate the development of advanced SDA technologies and methodologies that can overcome these limitations.

The Rayleigh criterion governs the limit to separate nearby sources in classical optics. The Rayleigh limit sets the minimum distance for point sources to be resolved. In SDA, the angular separation of targets is often less than the Rayleigh limit due to the large distance between the targets and the camera. Hence, the ability to resolve sources within the Rayleigh limit holds paramount importance in SDA \cite{choi2023telescope}. One of the earliest initiatives to do so is the Lincoln Near Earth Asteroid Research (LINEAR) program \cite{stokes2000lincoln}. LINEAR tried to overcome the Rayleigh limit and resolve sources by generating a binary map with a Binary Hypothesis Test (BHT) followed by a multi-hypothesis test (MHT) to filter out outliers \cite{zingarelli2014improving}. Despite these efforts by the LINEAR program, the detection of space targets usually requires complex procedures such as delicate background removal while preserving the desired signal \cite{li2021resolution}. In addition to these complexities, the task is further complicated in SDA where long exposure times cannot be guaranteed due to factors such as atmospheric turbulence and target motion \cite{Bao:21}, adding another layer of difficulty in resolving point sources.

These days, machine learning and deep learning approaches have received attention due to the dramatic development of graphics processing units (GPUs). However, these approaches often require striking features to detect targets like the long trajectory of the object \cite{de2022real}. Subdued object characteristics can hinder the detection performance \cite{jia2020detection,li2022bsc}. In addition, most deep learning and machine learning approaches require extensive pre-training which can be both time-consuming and computationally expensive. This resource-intensive process is particularly challenging when dealing with large-scale datasets or intricate models. Moreover, such methods often require vast amounts of training data to achieve optimal performance. However, in most cases in SDA, pre-labeled data is scarce or unavailable, prompting researchers to use artificially generated data to train the network \cite{9771446}.

The present authors proposed quantum-accelerated imaging (QAI) \cite{Bao:21} using an adaptive strategy. The technique successfully estimates incoherent sources within the sub-Rayleigh region, demonstrating an acceleration in measurement time over classical direct imaging without using prior information like known source number or equal brightness. Before the introduction of QAI, papers relied on one or more a priori information like equal brightness \cite{lupo2016ultimate}, the number of stars being known \cite{Paur:16,Yang:16}, and/or known centroids \cite{rehacek2017optimal} which often brings significant performance differences \cite{lee2022quantum}. In addition, QAI incorporates adaptive modes to maximize the quantum Fisher information rather than using a fixed modal basis. By doing so, with zero a priori information, QAI requires 10 to 100 times less photons, i.e., the measurement is accelerated in QAI compared to classical direct imaging. This acceleration leads a substantial reduction in the required number of photons, thereby enhancing the efficiency of the imaging process. 

\begin{figure}[h!]
    \centering
    \includegraphics[width=0.8\textwidth]{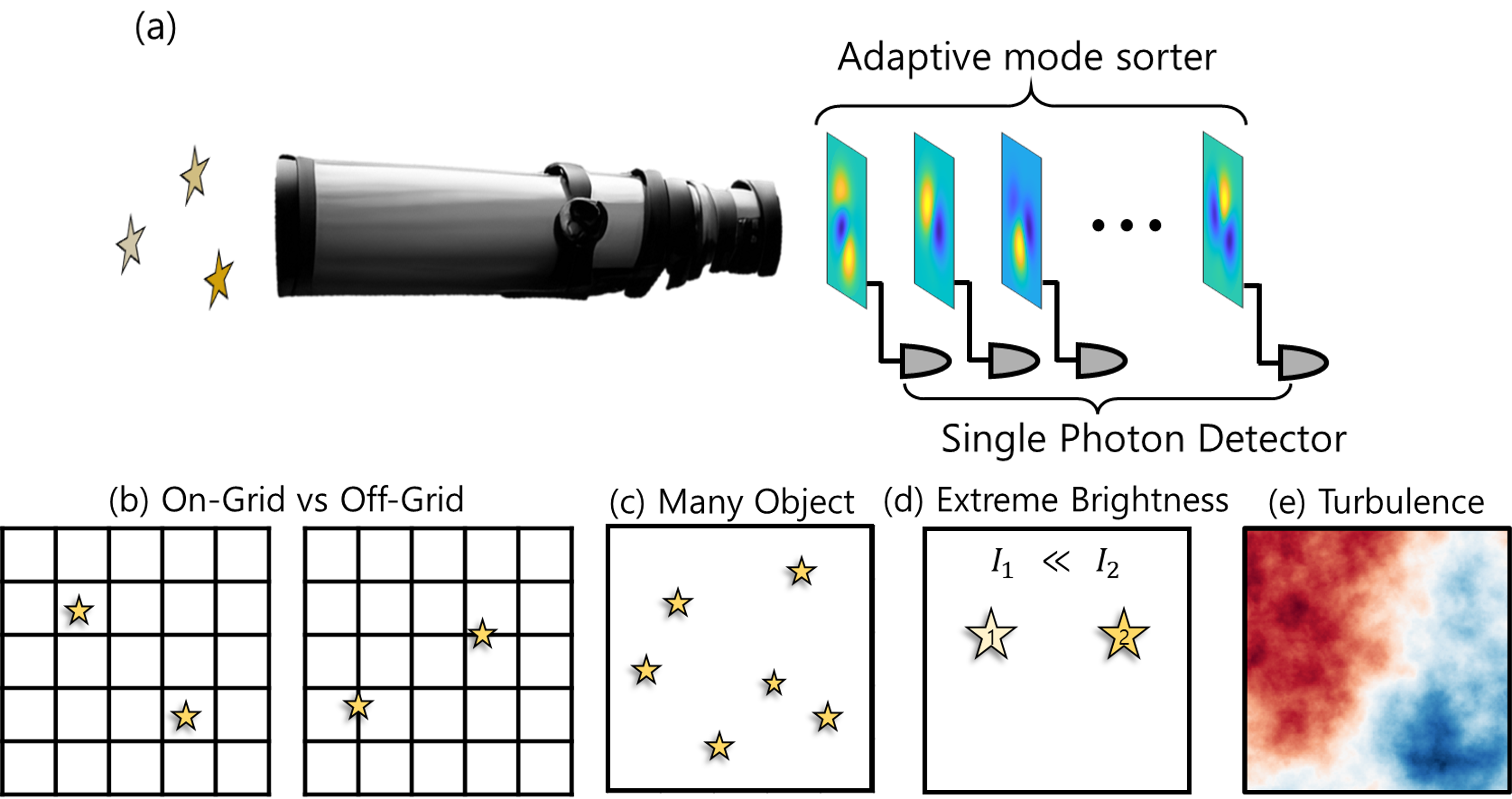}
    \caption{(a) Schematic of a mode sorting-based estimation. (b)-(e) Extreme constellation problem for space domain awareness (SDA). (b) On-grid and off-grid problems. (c) Many-object problem. (d) Extreme brightness ratio. (e) Atmospheric turbulence}
    \label{fig1}
\end{figure}

In Ref.\cite{Bao:21}, the study demonstrated Quantum Accelerated Imaging (QAI) algorithms along with a few basic examples using angular and radial grids. However, for SDA constellation problems, we encounter multiple off-grid targets with extreme brightness contrast as well as atmospheric turbulence. In response, this paper further generalizes QAI to address three specific challenges associated with SDA problems. (1) On-grid vs off-grid (2) Many object problem (3) Extreme brightness. (refer to figure \ref{fig1}). The first challenge pertains to the on-grid vs off-grid problem. First, we start from a comparison between ground truth targets located on-grid, using each pixel as a possible location, and targets located off-grid so that ground truth location can be anywhere within the image plane (see figure \ref{fig1}(b)). Here, we exhibit the performance of QAI where two targets in the scene are non-uniform and non-extreme brightness ratio. We then move on to more complex scenarios. These include many-object problem, such as scenes with 3 or more targets with a non-extreme brightness ratio (see figure \ref{fig1}(c)). Subsequently, we further develop many-object problem with an extreme brightness ratio. Explicitly, the brightest target being 100 or 1,000 times brighter than the faintest target scenarios are explored (see figure \ref{fig1}(d)). Each challenge has been thoroughly tested with at least 1000 Monte Carlo simulations. Overall, in all three challenges, QAI shows better performance than any other state-of-the-art method. In particular, QAI shows up to a 3x higher chance of correctly detecting the source and 8x better in brightness estimation. Moreover, in contrast to other studies that require 10,000 or more photons per target for successful detection \cite{tsang2017subdiffraction}, our approach is more stringent, limiting the number of photons to 1,000 per target. Therefore, with superior performance in every tested aspect, QAI stands out as the most effective solution for SDA problems.

Atmospheric turbulence is another key challenge in SDA \cite{roggemann2018imaging}. When light travels through the turbulence, random variations in air density lead to the scattering and refraction of light. As a result, the resulting image gets distorted and blurred, reducing the optical resolution and making it even more difficult to resolve the nearby point sources. Techniques such as adaptive optics \cite{roddier1999adaptive}, which use deformable mirrors and wavefront sensors to correct for the distortions caused by atmospheric turbulence, are employed to mitigate this effect. However, the use of a wavefront sensor (usually Shack-Hartmann wavefront sensor) poses a limitation on spatial resolution due to the size of microlenses and requires a significant number of photons for successful measurement. Therefore, we utilize QAI in the presence of atmospheric turbulence for detecting and localizing the point sources. Here, we followed Noll's approach \cite{noll1976zernike} for turbulence generation. By incorporating atmospheric turbulence effects into our estimations, following the discretization approach of Ref. \cite{Bao:21}, QAI shows a great ability to resolve nearby sources even with the presence of turbulence. The comprehensive simulations conducted in this study are illustrated in the second row of figure \ref{fig1}. These simulations contribute significantly to our understanding of the practicality and potential of QAI in addressing SDA problems.

We have organized the paper in the following way. First, we provide a review of QAI and how it can detect and localize the resident space objects. Following this, we commence with a binary star case, starting with the on-grid scenario. In this context, we evaluate the performance of QAI and compare it with other state-of-the-art methods. Subsequently, we extend this comparison to off-grid scenarios, examining how the performance shifts when the binary stars are not aligned with the grid. Subsequently, we assess the performance of QAI for the many-object problem, where more than binary stars are located within a sub-Rayleigh regime. Finally, we evaluate the extreme brightness ratio cases (1:100 and 1:1000). Through these evaluations, we aim to confirm the advantages of QAI in various astronomical scenarios and to demonstrate its potential as a powerful tool in Space Domain Awareness (SDA) and other related fields.

\section{Quantum-Accelerated Imaging (QAI)}

\begin{figure}[h!]
    \centering
    \includegraphics[width=\textwidth]{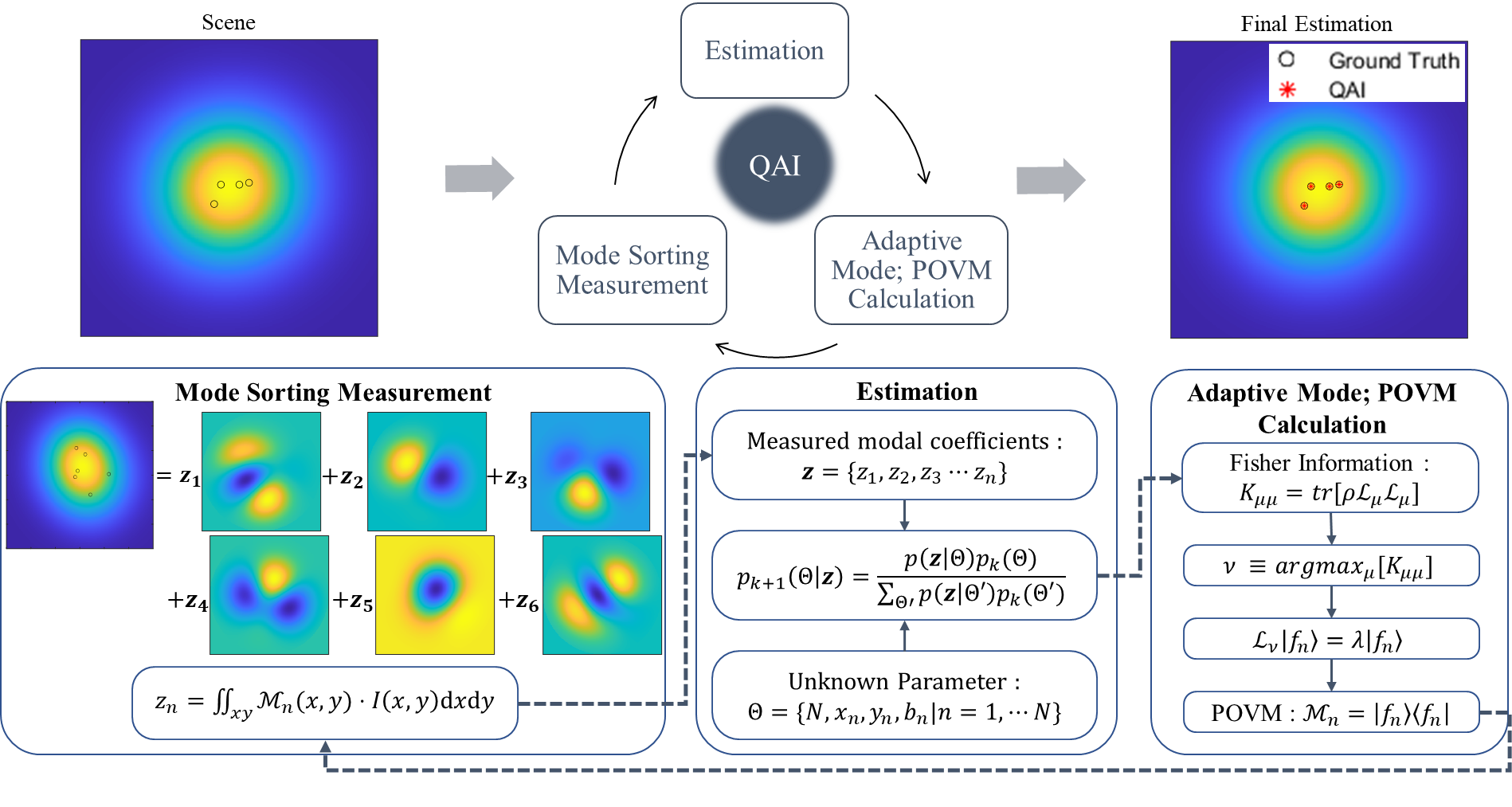}
    \caption{Schematics of Quantum-Accelerated Imaging (QAI). QAI is an iterative algorithm that consists of three stages: mode sorting measurement, adaptive mode; POVM calculation, and estimation. In the mode sorting measurement stage, the scene is projected onto the computed modes, and the corresponding modal coefficients are measured. Next, in the estimation stage, calculate the parameter set to maximize the likelihood of parameters given measurement data. Lastly, in the adaptive mode; POVM calculation stage, we calculate a new adaptive mode; POVM based on the previous estimation, maximizing the Quantum Fisher Information (QFI). For the initial iteration, we adopt Zernike modes for the adaptive modes 
}
    \label{fig2}
\end{figure}

QAI \cite{Bao:21} is an iterative method that consists of three stages: mode sorting measurement, adaptive mode calculation, and estimation, as illustrated in figure \ref{fig2}. In the mode sorting measurement stage, QAI measures the coefficient corresponding to each mode using an adaptive mode sorter and single photon detector as depicted in figure \ref{fig1}. Here, for a given scene $I(x,y)$ the adaptive mode sorting can be denoted as
\begin{equation}
    z_n=\iint_{xy} M_n(x,y) \cdot I(x,y)dxdy
\end{equation}
where, $M_n(x,y)$ is an n-th mode which is calculated in the adaptive mode calculation stage and $z_n$ is the corresponding coefficient. Next, in the estimation stage, we calculate the probability for estimation following photon statistics. With the calculated probability, the parameter set with the largest probability will be our estimation for each iteration. Here, the parameters of interest are the number of sources in the scene and the position and brightness of each source. A detailed explanation of the estimation can be found in the appendix. Lastly, in the adaptive mode calculation stage, the preceding estimation determines the subsequent adaptive mode that maximizes the Quantum Fisher Information (QFI) as QFI is always greater than or equal to classical Fisher information. Please refer to the appendix for details on calculating the modes. For the initial measurement and estimation, we adopt the first 10 Zernike modes. This iteration continues until it reaches the photon budget. Upon reaching this budget, the iterative process is halted, providing the final estimation result.

\section{Numerical Simulations}
\subsection{Simulation Settings}
Next, we present the results of numerical simulations of QAI and state-of-the-art direct imaging methods. Figure \ref{fig2} shows an example figure of the input scene of incoherent sources blurred by a circular aperture with an Airy disk point spread function (PSF). Without losing generality, we adopted a 2D Gaussian function as a PSF, with the width determined by the aperture size. Here, we define a Rayleigh length ($\mathrm{rl}$) as 
\begin{equation}
    \mathrm{rl} = \frac{1.22}{4} \pi \frac{f}{d} \sigma
\end{equation}
where, $f$ is focal length, $d$ is a lens diameter and $\sigma$ is standard deviation of Gaussian function. Assuming $1m$ focal length and $0.1m$ lens diameter, Rayleigh length ($\mathrm{rl}$) is
\begin{equation}
    \mathrm{rl} = 9.577 \sigma \approx 10\sigma.
\end{equation}
The parameters of interest include the number of stars in the scene ($N$), their positions ($x_n,y_n$), and brightness ($b_n$), which form the search space $\Theta=\{N,x_n,y_n,b_n | n=1,\cdots, N\}$. 
Here, we set the search range for the number of stars, $N$, to exceed the number of stars in the scene to ensure a fair estimation. The imaging size is $64\times64$. However, the computational cost of searching the entire area is prohibitive. Therefore, we limit our search to a smaller central region since the PSF may not be located in the outer region. Here, under the consideration of shot noise, we operate under an extreme photon-starved regime, utilizing only 1000 photons per star for the estimation, unless otherwise specified. 

To ensure robustness and reduce the impact of random variation, we conducted a large number of Monte Carlo simulations - at least 1000 simulations for each scenario, except for the 6-star case. For the performance evaluation, we compare the accuracy of star number estimation in the scene for performance evaluation. Subsequently, for the successful number estimation cases, we examine the accuracy of brightness and position estimation. For position estimation, we adopt root mean square error, as shown below:
\begin{equation}
    ||Error||  =   \sqrt{ \frac{\sum_{i=1}^n \left[ \left( x_i -\hat{x}_i \right) ^2 + \left( y_i -\hat{y}_i \right) ^2 \right]}{N} }.
\end{equation}
Here, $(x_i,y_i)$ is the ground truth location of i-th source and $(\hat{x}_i,\hat{y}_i)$ is the estimated position.

\subsection{Simulation Results}

\subsubsection{On-grid vs Off-grid} \hspace*{\fill} \\

\begin{figure}[htb!]
    \centering
    \includegraphics[width=0.9\textwidth]{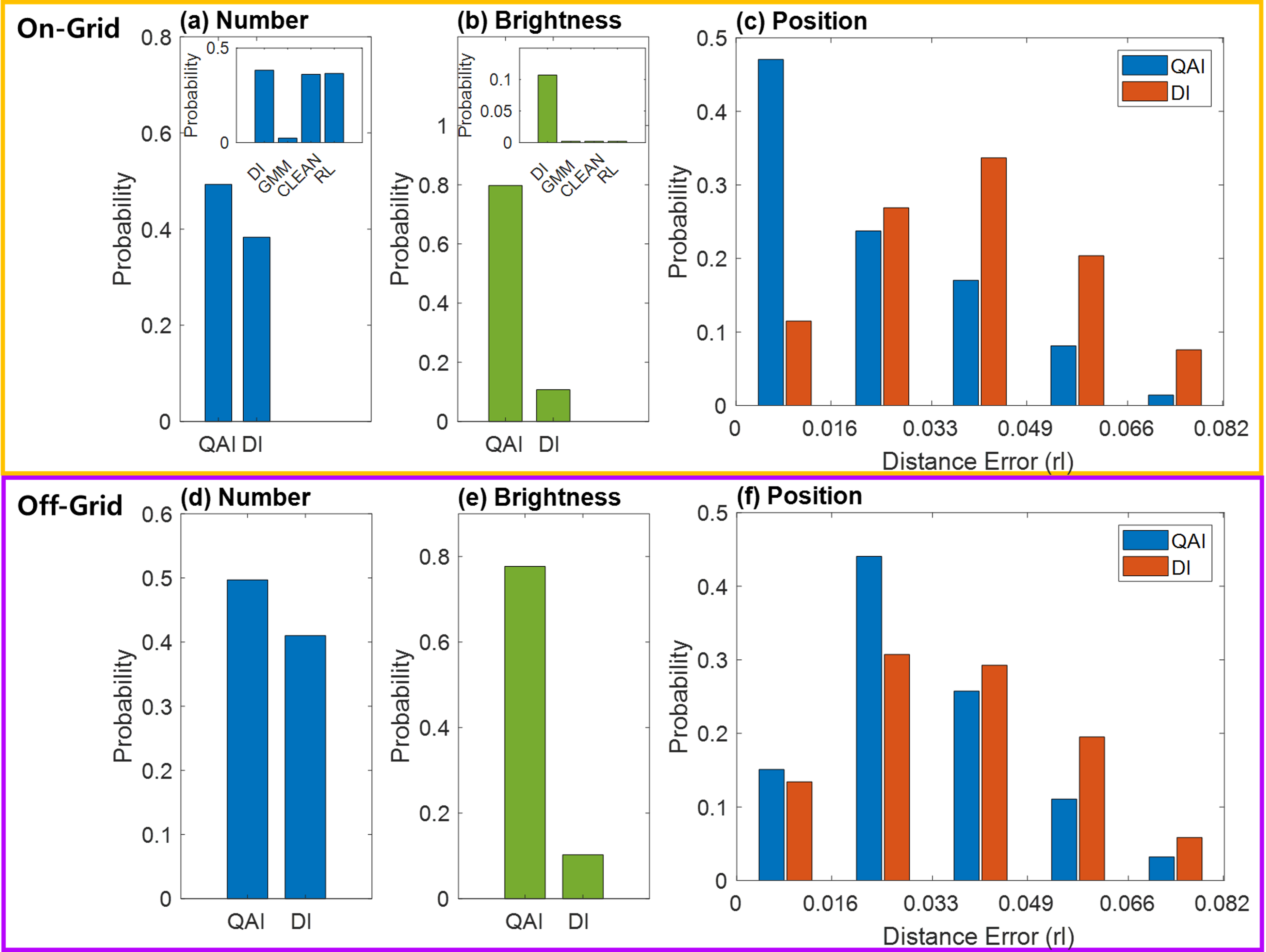}
    \caption{QAI performance demonstration compared to direct imaging (DI) using L-2 cost in on-grid (a)-(c) and off-grid (d)-(f) star scenarios. Insets compare DI with GMM, CLEAN, and RL. In all cases, QAI outperforms DI, and QAI and DI show similar performance between on-grid and off-grid scenarios.
    (a) Probability for correct number estimation of the stars in the scene. (b) Probability for correct brightness estimation. (c) Position estimation error distribution. (d) The probability for correct number estimation of stars located off-grid. (e) Probability for correct brightness estimation of stars located off-grid. (f) Distribution of position estimation errors for QAI and DI for stars located off-grid.}
    \label{fig3}

\end{figure}

We start by considering two stars in the scene with brightness levels discretized by 1, 2, or 3. As the pixel discretization issue is a key limiting factor of centroid estimation, we compare the performance of the on-grid (figure \ref{fig3}(a)-(c)) and off-grid (figure \ref{fig3}(a)-(c)) case in this section. For the on-grid case, we considered the center of each pixel as the ground truth centroid location for the stars. In contrast, for the off-grid scenario, we assumed that the ground truth star centroid location could be anywhere, while still being estimated using on-grid locations. The performance of comparison methods (DI, GMM, CLEAN, RL) is also presented in the inset of each figure. Please refer to the appendix for a detailed explanation of comparison methods.    

Here, we discuss the probability of correct number and brightness estimation and position estimation error distribution. Figure \ref{fig3}(a) shows the probability for the correct number estimations for QAI and DI. QAI shows around 30\% improvement in number estimation accuracy. Figure \ref{fig3}(b) presents the probability for the correct brightness estimation. QAI significantly outperforms our DI showing an 8 times larger probability to correctly estimate brightness. Lastly, figure \ref{fig3}(c) displays the distribution of position estimation errors, where QAI surpasses DI in location estimation as well. The inset figure of figure \ref{fig3} (a) and (b) shows the performance of our DI with other state-of-the-art methods (GMM, CLEAN, RL) where our DI shows the best performance among the methods. The mean position estimation error values for other direct imaging methods are tabulated in table \ref{tab1} with our DI showing the lowest mean position estimation error. Among the comparison methods, our DI shows the best performance in every aspect. Thus, for the following results, we will compare the performance of QAI and our DI method. 

\begin{table}[hbt!]
    \centering
    \begin{tabular}{c|c|c|l}
    \multicolumn{1}{l|}{Method} & Error $\pm$  $\sigma$ & \multicolumn{1}{l|}{Method} & \multicolumn{1}{c}{Error $\pm$  $\sigma$} \\ \hline
    DI                          & 0.0425 $\pm$  0.0158  & CLEAN                       & 0.0643 $\pm$  0.0277                      \\
    GMM                         & 0.0492 $\pm$  0.0148  & RL                          & 0.0521 $\pm$  0.0178                     
    \end{tabular}
    \caption{Mean error for direct imaging methods. Our DI method shows the lowest mean error.}
    \label{tab1}
\end{table}



Next is the performance of off-grid scenarios (figure \ref{fig3}(d)-(f)). In all the three parameter estimations shown, we can clearly see that the QAI shows a better estimation performance than DI. More importantly, comparing the performance of on-grid and off-grid scenarios, both methods show remarkable consistency in estimation. Therefore, for the following result, we will focus on the on-grid estimations otherwise mentioned. Please note that, for position estimation in the off-grid scenario, we adopt pixel centroid for localization. In contrast, the ground truth location is set to off-grid, resulting in some degree of small error.

\subsubsection{Many-object problem} \hspace*{\fill} \\

\begin{figure}[ht!]
    \centering
    \includegraphics[width=0.9\textwidth]{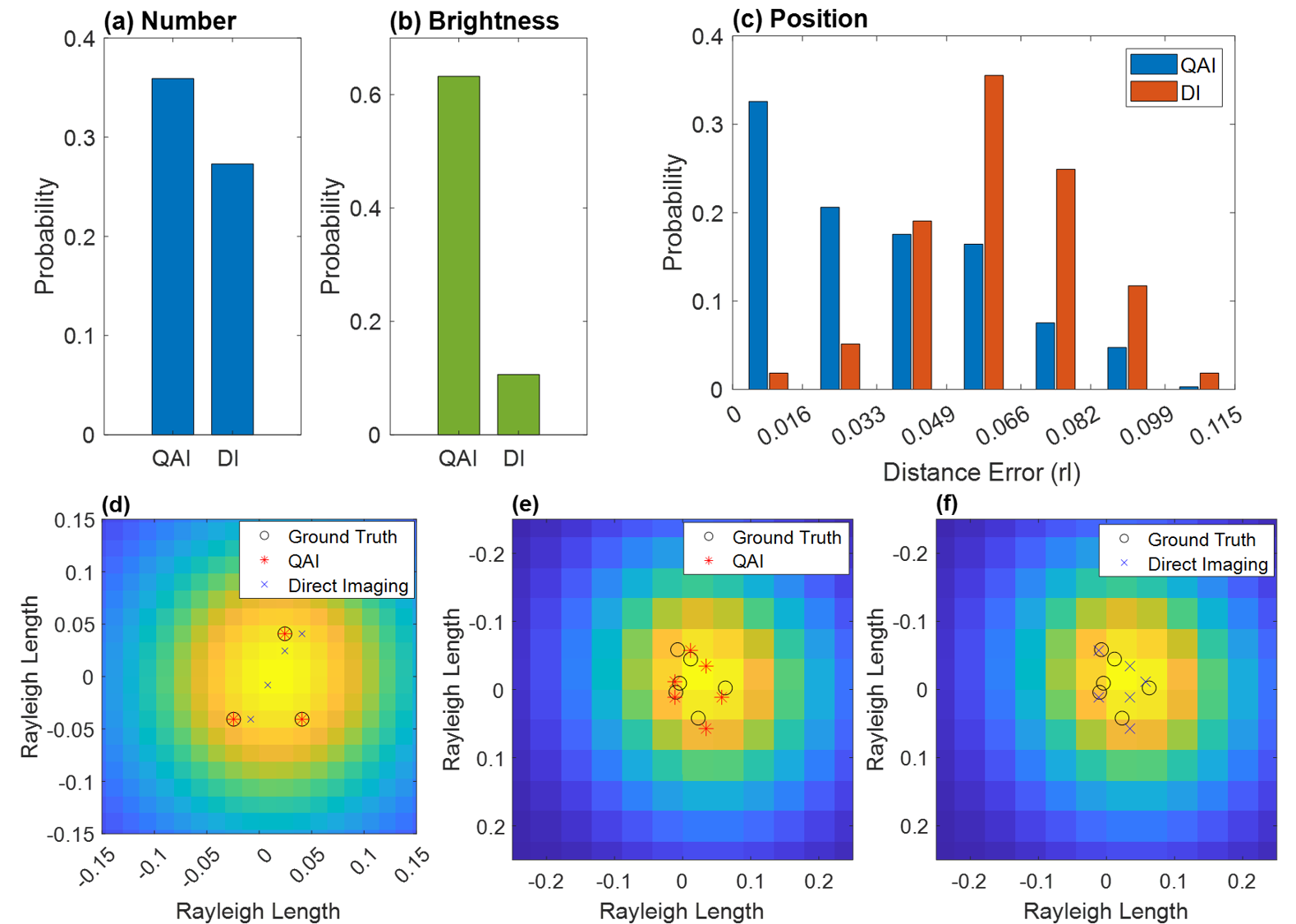}
    \caption{Demonstration of QAI for many-object problem. QAI outperforms DI in all cases. (a) Probability of correct number estimation for a 3-star scenario. (b) Probability of correct brightness estimation for a 3-star scenario. (c) Position estimation error distribution for QAI and DI. (d) Example figure for QAI and DI performance in a 3-star scenario. (e)-(f) Example figures for QAI and DI for the same scene for a 6-star scenario with off-grid ground truth star locations. (e) QAI successfully estimated the scene. (f) DI failed to correctly guess the number of stars and also showed inaccurate location estimation.
}
    \label{fig4}
\end{figure}


Next, we expand our investigation from the binary star case to the many-object problem. The estimation performance for a 3-star scenario is shown in figure \ref{fig4}(a)-(c). Our proposed method, QAI, consistently outperforms DI for all parameter estimations, including number, brightness, and position estimation. As demonstrated in figure \ref{fig4}(d), QAI achieves correct estimation of the scene, while DI fails to accurately estimate the number of stars in the scene.

Another example figure for many-object problem can be found in figure \ref{fig4}(e)-(f), where we analyze the performance of the methods for the 6-star scenario with ground truth stars located off-grid. QAI shows satisfying estimation performance in the mentioned complex scene whereas DI tends to fail to correctly estimate the number of stars in the scene. Therefore, this underscores the supremacy of QAI in many-object problem. Here, due to computational limits, we were not able to run enough cases to generate statistics for the 6-star case.

\subsubsection{Extreme brightness} \hspace*{\fill} \\

\begin{figure}[ht!]
    \centering
    \includegraphics[width=0.9\textwidth]{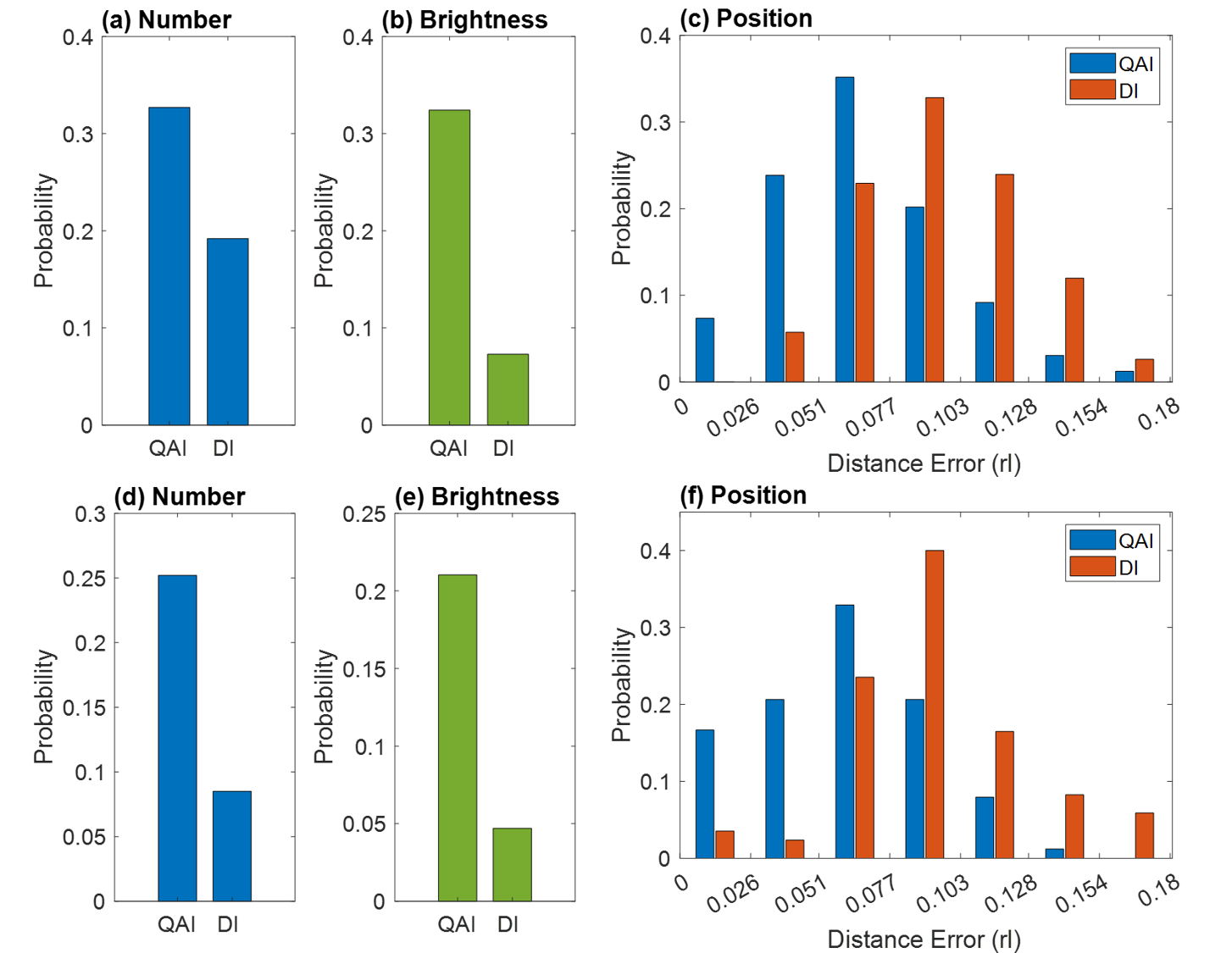}
    \caption{Demonstration of QAI for an extreme brightness cases; 100 or 1000 brightness ratio. In all cases, QAI outperforms DI. (a) Probability for correct number estimation of stars with 100 times brightness ratio. (b) Correct brightness estimation of stars with 100 times brightness ratio. (c) Position estimation error distribution with 100 times brightness ratio. (d) Probability for correct number estimation of stars with 1000 times brightness ratio. (e) Correct brightness estimation of stars with 1000 times brightness ratio. (f) Position estimation error distribution with 1000 times brightness ratio. 
}
    \label{fig5}
\end{figure}

To further demonstrate the effectiveness of QAI, we expand our investigation to the many-object problem with extreme brightness ratio scenarios, as shown in figure \ref{fig5}. In figure \ref{fig5}, the top row figures ((a)-(c)) illustrate a 100 times brightness ratio and the bottom row figures ((d)-(f)) illustrate 1000 times brightness ratio scenarios. A photon budget of 2500 photons per star is utilized for estimation in the former case and 15000 photons for the latter case. Our QAI method consistently outperforms DI in both extreme brightness ratios. Notably, QAI shows a remarkable ability to handle extreme brightness ratios in a photon-starved situation. Although there is a slight performance degradation compared to previous sections, considering the sparse amount of photons we collect from the dimmest star due to extreme brightness contrast, QAI is performing well in estimation. Overall, these results highlight the capability of our method to handle challenging scenarios in a photon-limited environment, emphasizing the significant advantages of QAI over DI.

\subsubsection{Atmospheric Turbulence} \hspace*{\fill} \\

\begin{figure}[h!]
    \centering
    \includegraphics[width=0.8\textwidth]{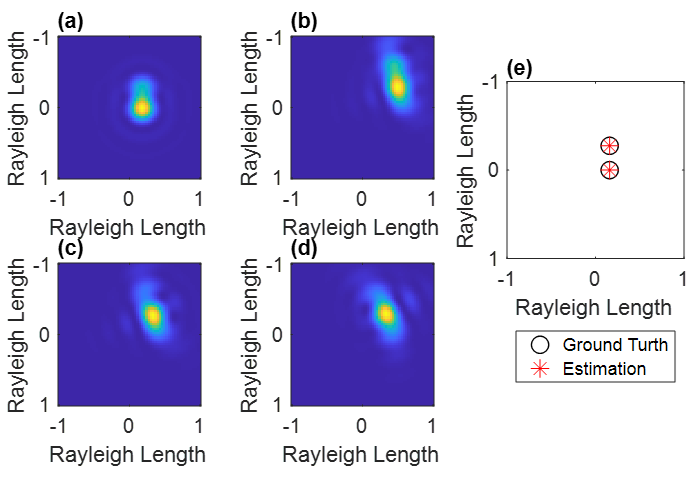}
    \caption{QAI performance demonstration in the presence of atmospheric turbulence. QAI successfully estimates the turbulence and the scene characteristics with only 249 photons.  (a) Ground truth scene without turbulence. (b) Ground truth scene with turbulence. (c) Ground truth turbulence point spread function (PSF) (d) Estimated turbulence PSF (e) Point source estimation result with QAI accurately estimates the position.}
    \label{fig6}

\end{figure}

Lastly, we study the case with the presence of atmospheric turbulence. Inspired by Noll's approach \cite{noll1976zernike}, we simulated and estimated the turbulence using Zernike polynomials \cite{von1934beugungstheorie}. The turbulence profile $\phi(r,\rho)$ can be denoted as
\begin{equation}
    \phi(r,\rho)=\sum_j a_j Z_j(r,\rho)
\end{equation}
where, $Z_j(r,\rho)$ is j-th Zernike polynomial and $a_j$ is the corresponding coefficient for j-th Zernike polynomial which is randomly drawn \cite{noll1976zernike}. Here, due to limited computational power, we limit the number of polynomials to 8 Zernike modes, and coefficients were discretized into two values, -1, and 1. Apart from the original QAI, to estimate a scene with turbulence, coefficients of the turbulence profile and the scene parameters are estimated concurrently so that QAI can successfully estimate the scene in the presence of turbulence. The result for the estimation is presented in figure \ref{fig6}. The ground truth scene without the effect of turbulence is depicted in figure \ref{fig6}(a) and the scene after passing the atmospheric layer is depicted in figure \ref{fig6}(b). The ground truth turbulence PSF can be found in figure \ref{fig6}(c). Comparing the estimated turbulence PSF, figure \ref{fig6}(d), with figure \ref{fig6}(c), we can see that QAI correctly estimates the turbulence. The source estimation result is described in figure \ref{fig6}(e) and we can observe QAI correctly estimates the scene even with the presence of atmospheric turbulence. Here we adopted two radial grids and six angular grids to reduce the computation. Here, we modify the exit criterion of QAI so that it runs until the result converges and gives 3 same estimation results for each iteration which results in a total of only 249 photons being used for estimation. Considering the presence of turbulence and the low source separation, the number of photons required for successful estimation is very low though we adopted the angular and radial grids. Due to the significant computational challenges, a more comprehensive exploration of these issues will be undertaken in our future work.

\section{Conclusion}
In this study, we tackled SDA challenges with QAI and showed improved performance against existing state-of-the-art methods. Our results demonstrated the superiority of QAI in SDA, like detecting a dim space target in a photon-starved situation. From binary stars with one order of magnitude brightness ratio to many-object scenarios with 1,000 times brightness ratio, QAI consistently outperforms other methods. Especially, as the scene gets more complex, QAI shows consistent performance whereas DI tends to show performance degradation. Lastly, we present an initial successful result of QAI estimation of the scene with turbulence. Overall, our findings highlight the significant potential and superiority of QAI for high-precision astronomical imaging in space domain awareness. However, due to a lack of computational resources, we were only able to run with some limited-scaled cases, especially for the turbulence scenario. Therefore, with a proper optimization to find maximum probability like the gradient descent approach, we expect to test QAI for more advanced turbulence scenarios.

\section*{Acknowledgements}
This work was partially supported by the Army Research Office (W911NF-21-1-0287) and Defense Advanced Research Projects Agency (DARPA). The authors thank Johns Hopkins Applied Physics Laboratory for providing the code for CLEAN and Richardson Lucy Deconvolution. The authors also wish to thank Seungman Choi, Peter Menart, and Andrew Schramka for reviewing the manuscript.

\section*{References}

\bibliographystyle{unsrt}
\bibliography{QAI}


\newpage
\appendix

\section{QAI}

For the estimation of QAI, we start by deriving a governing equation for calculating the coefficient based on a number of sources ($N$) and each position($x_n,y_n$) and brightness($b_n$). Here, estimated scene intensity,$\hat{I}(x,y)$ is 
\begin{equation}
    \hat{I}(x,y)=\sum_n b_n P(x-x_n,y-y_n)
\end{equation}
and, thereby estimated coefficient is computed as
\begin{equation}
    \hat{z}_n=\iint_{xy} M_n(x,y) \cdot \hat{I}(x,y)dxdy.
\end{equation}
Here, note that the adaptive mode $M_n(x,y)$ is the same as the one used in the measurement stage which is calculated in the adaptive mode calculation stage.  
Next, the probability of measuring each coefficient is calculated using the Poisson distribution. The probability of measuring $z_n$ photon, $p(z_n)$, assuming $\hat{z}_n$ photon estimated is 
\begin{equation}
    p(z_n) = \frac{{\hat{z}_n}^{z_n} \  e^{-\hat{z}_n}}{z_n!}.
\end{equation}
Here, as each mode projection is an independent event, by taking a product of each probability, we can compute the total probability 
\begin{equation}
    p(\textbf{z})=\prod_n p(z_n)
\end{equation}
where, $\textbf{z}=\{z_1,z_2,z_3\cdots z_n\}$ is a measured coefficients set. Considering QAI is an iterative method, for better estimation, we adopt Bayesian inference. For a given prior probability mass function (PMF), $p_k(\Theta)$, in the parameter space, $\Theta$, and observed new data $\mathbf{z}$ in the mode sorting measurement stage for $k$-th iteration, the posterior PMF is denoted as
\begin{equation}
    p_{k+1}(\Theta|\textbf{z}) = \frac{p(\textbf{z}|\Theta)p_k(\Theta)}{\sum_{\Theta'} p(\textbf{z}|\Theta')p_k(\Theta')}.
\end{equation}

Next, we address how to compute adaptive modes by maximizing quantum Fisher information. The parameters used are defined in the Table \ref{tab2}.
\begin{table}[h!]
    \centering
\begin{tabular}{|l|l|}
\hline
Parameter           & Description                                     \\ \hline
$\rho$              & density operator                                \\ \hline
$b_n$               & brightness of $n_\mathrm{th}$ star                         \\ \hline
$\calL$             & symmetric logarithmic derivative (SLD) operator \\ \hline
$\ket{e_n}$         & eigenbases of $\rho$                            \\ \hline
$D_n$               & eigenvalue of $\rho$                            \\ \hline
$K_\mu$             & quantum Fisher information                      \\ \hline
$\calM_n$           & positive operator-valued measures (POVMs)       \\ \hline
$\Theta$            & parameter estimation set                        \\ \hline
$z$                 & observed new data                               \\ \hline
$p_k(\Theta)$       & probability mass function at $k_\mathrm{th}$ iteration     \\ \hline
\end{tabular}
    \caption{Parameter Definition}
    \label{tab2}
\end{table}
The quantum state of light, ignoring two or more photons arriving at the same time, can be described as 
\begin{equation}
    \rho_\mathrm{state} = (1-\epsilon)\rho_0 + \epsilon\rho_1,
\end{equation}
where, $\epsilon$ denotes average number of photons, $\rho_0$ is the zero-photon state, $\rho_1$ is the one-photon state. Here, we consider a weak-signal situation so that 
\begin{equation}
    \epsilon << 1.
\end{equation}
One-photon state, $\rho_1$, further denote as $\rho$, for N incoherent point sources can be described as 
\begin{equation}
    \rho = \sum_{n=1}^N b_n\ket{\phi_n}\bra{\phi_n},
\end{equation}
where the point spread function (PSF) of the $n_\mathrm{th}$ star on the image plane is
\begin{equation}
    |\proj{x,y}{\phi_n}|^2 = \Psi(x-x_n,y-y_n,\sigma)
\end{equation}
where, $b_n$ is the brightness of each point source normalized so that $\sum_N b_n = 1$.
In the sub-Rayleigh regime, as the sources are located nearby, overlapping each other, $\proj{\phi_m}{\phi_n}\neq 0$ for $m\neq n$.
Here, the density operator has an $N\times N$ matrix form.
For a diagonalized density matrix of unknown N incoherent point sources and an estimated parameter $\mu\in \{ x_n,y_n,b_n \}$, the symmetric logarithmic derivative (SLD) operator $\calL_\mu$ can be denoted as  
\begin{equation} 
    2\p_\mu \rho = \calL_\mu \rho + \rho \calL_\mu.
\end{equation}
Quantum Fisher information can be defined as $Tr(\rho \calL_\mu \calL_\mu)$.
The SLD operator $\calL_\mu$ generally can be expanded by the eigenbases of $\rho$, $\{ \ket{e_n} \}$, and their derivatives with respect to $\mu$. Its sub-matrix in the $\{ \ket{e_n} \}$ space, however, proves efficient in finding the near-optimal measurement. The sub-matrix solution for the SLD operator $\calL_\mu$ is 
\begin{equation}
    \calL_\mu = \sum_{mn}\frac{2}{D_m+D_n}\bra{e_m}\p_\mu\rho \ket{e_n}\cdot \ket{e_m}\bra{e_n},
\end{equation}
where $D_n$ is the corresponding eigenvalues of $\ket{e_n}$ computed by singular value decomposition (SVD). 
Here, the quantum Fisher Information for a single parameter per a detected photon is defined using the SLD operator as
\begin{equation}
    K_{\mu\mu} = \tr[\rho \calL_\mu \calL_\mu].
\end{equation}
Here, we define, $\nu$ to be the parameter that maximizes the quantum Fisher information, $\nu \equiv \mathrm{argmax}_\mu[K_{\mu\mu}]$.
The eigenbasis of $\calL_\nu$ to be $\ket{f_n}$, where, $\ket{f_n}$ provides a set of near-optimal and adaptive bases for estimating the parameter $\nu$. Here, positive operator-valued measures (POVMs) can be denoted as
\begin{equation}
    \calM_n = \ket{f_n}\bra{f_n},\quad n=1,\cdots,N.
\end{equation}

In the previous paper, Ref.\cite{Bao:21}, we adopted radial and angular grids for possible star locations due to computational limits. However, to explore more general real-world suited cases we removed the grid and searched all possible star locations in this study. Here, due to the fundamental principle of quantum metrology that maximizing QFI always gives sub-optimal solutions, QAI is expected to outperform classical estimation methods.

\section{Baseline Direct Imaging}

For the performance comparison, we first analyze the three state-of-the-art, widely adopted in SDA, direct imaging methods that utilize the classical domain intensity profile of an imaging system; Gaussian Mixture Model \cite{reynolds2009gaussian}, CLEAN algorithm \cite{hogbom1974aperture}, and Richardson Lucy deconvolution algorithm \cite{richardson1972bayesian,lucy1974iterative}. We further compare them with our highly optimized L2 cost function based direct imaging method. 

A Gaussian Mixture Model (GMM) is a probabilistic model representing a probability distribution as a combination of multiple Gaussian distributions. GMM adopts an expectation-maximization (EM) algorithm to fit given data into Gaussian distributions. In the expectation step (E-step), calculate each data point belongs to a certain cluster. Then, in the maximization step (M-step), update the parameters like centroid x and y location to maximize the probability. These two steps repeat so that the probability can be maximized and give the final estimation output. 

The CLEAN algorithm works by iteratively subtracting a model of the point spread function (PSF) from the image. Starting with the highest intensity point, subtract the PSF shifted to the highest intensity point while storing the highest intensity point. After, from the subtracted image, repeat the subtraction process until it reaches the threshold. After, using the stored peak points, with post-processing methods like density-based clustering, the final estimation for a number of sources and their location and brightness can be computed. This process effectively removes the effects of PSF blur allowing estimation of the centroid location of the sources.

The Richardson-Lucy deconvolution algorithm is a widely used iterative method for image deconvolution. It is commonly used in microscopy and astronomy to enhance the quality of images degraded by blur or noise. The algorithm uses a model of the point spread function (PSF) to estimate the original image from the blurred image. It iteratively refines the estimate by alternating between the forward model and a backward update step that accounts for the PSF. The algorithm converges to an estimate of the original image that has been deconvolved with the PSF. After deconvolution, sources with intensity above a threshold can be assumed to be centroids of photon sources. 

For our L2 cost function-based direct imaging method, the method calculates the cost as follows
\begin{equation}
    Cost = \sum_{xy}{\left(I_\mathrm{meas}(x,y) - I_{i}(x,y)\right)^2}
\end{equation}
where $I_\mathrm{meas}$ is a measured scene from the detector considering shot noise and $I_i(x,y)$ is i-th possible estimated scene. After calculating the cost, we choose the estimated scene that gives the lowest cost as the final estimation. This is because the cost function calculates the error for each possible scene, and selecting the scene with the lowest error ensures that we choose the most accurate estimation. For the rest of the paper, we will call our direct imaging method DI.

\end{document}